\documentclass[runningheads]{llncs}
\pdfoutput=1
\usepackage{graphicx}
\usepackage{amsmath}
\usepackage{gensymb}
\usepackage{subfloat}
\usepackage{slashbox}
\usepackage{placeins}
\usepackage[strings]{underscore}
\usepackage{etoolbox}
\usepackage{booktabs}
\apptocmd{\sloppy}{\hbadness 10000\relax}{}{}

\usepackage{cleveref}
\usepackage{acronym}
\acrodef{OD}[OD]{optic disc}
\acrodef{SLO}[SLO]{scanning laser ophthalmoscope}
\newacro{UWFoV-SLO}{ultra-wide field of view scanning laser ophthalmoscope}
\newacro{FoV}{field of view}
\newacro{FC}{fundus camera}
\newacro{CNN}{convolutional neural network}
\acrodef{AF}[AF]{autofluorescence}
\acrodef{PCA}[PCA]{principal component analysis}
\acrodef{SVM}[SVM]{support vector machine}
\acrodef{RG}[RG]{red-green}
\acrodef{ES}[ES]{eyesteered}
\acrodef{CP}[CP]{central-pole}
\acrodef{DL}[DL]{deep learning}

\begin{document}
\title{Optic disc and fovea localisation in ultra-widefield scanning laser ophthalmoscope images captured in multiple modalities}
\titlerunning{Landmark localisation in retinal images}
% If the paper title is too long for the running head, you can set
% an abbreviated paper title here
%
\author{Peter R. Wakeford\inst{1,2} \thanks{Supported by the EPSRC Centre for Doctoral Training in Applied Photonics.}\and
Enrico Pellegrini\inst{2} \and
Gavin Robertson\inst{2} \and
Michael Verhoek\inst{2} \and
Alan D. Fleming\inst{2} \and
Jano van Hemert\inst{2} \and
Ik Siong Heng\inst{1}}
%

% PW  \orcidID{0000-0002-4547-0087}
% EP \orcidID{0000-0001-6913-1760}
% jvH \orcidID{0000-0003-0834-7079}
% ISH \orcidID{0000-0002-1977-0019}
\authorrunning{P. Wakeford et al.}
% First names are abbreviated in the running head.
% If there are more than two authors, 'et al.' is used.
%
\institute{University of Glasgow, Glasgow, Scotland, G12 8QQ \and
Optos (Nikon), Carnegie Campus, Dunfermline, Scotland, KY11 8GR
}
\maketitle              % typeset the header of the contribution

\begin{abstract}
We propose a convolutional neural network for localising the centres of the \ac{OD} and fovea in \ac{UWFoV-SLO} images of the retina. Images captured in both reflectance and autofluorescence (AF) modes, and central pole and eyesteered gazes, were used. The method achieved an \ac{OD} localisation accuracy of 99.4\% within one \ac{OD} radius, and fovea localisation accuracy of 99.1\% within one \ac{OD} radius on a test set comprising of 1790 images. The performance of fovea localisation in AF images was comparable to the variation between human annotators at this task. The laterality of the image (whether the image is of the left or right eye) was inferred from the \ac{OD} and fovea coordinates with an accuracy of 99.9\%.
\end{abstract}

\section{Introduction}
\acresetall
The \ac{OD} and fovea are two anatomical landmarks found in the retina. The \ac{OD} is where vasculature and nervous connections are made to the rest of body, and appears as a bright disc or oval in retinal images. The fovea, the centre of vision, is the area with the highest density of cone receptor cells, and is found on-axis to the lens of the eye.

Automating the location of these landmarks in retinal images is an important first step in the computer-aided diagnosis of many retinal diseases. For example, glaucoma can be graded by measuring morphology parameters of the \ac{OD} \cite{Tangelder2006}, and automation of this measurement requires the \ac{OD} to be located \cite{Haleem2013}. In the detection of diabetic retinopathy, it is often necessary to identify the \ac{OD} before detection of exudates \cite{Faust2012}.

\begin{figure}[!ht]
  {\includegraphics[width=\linewidth]{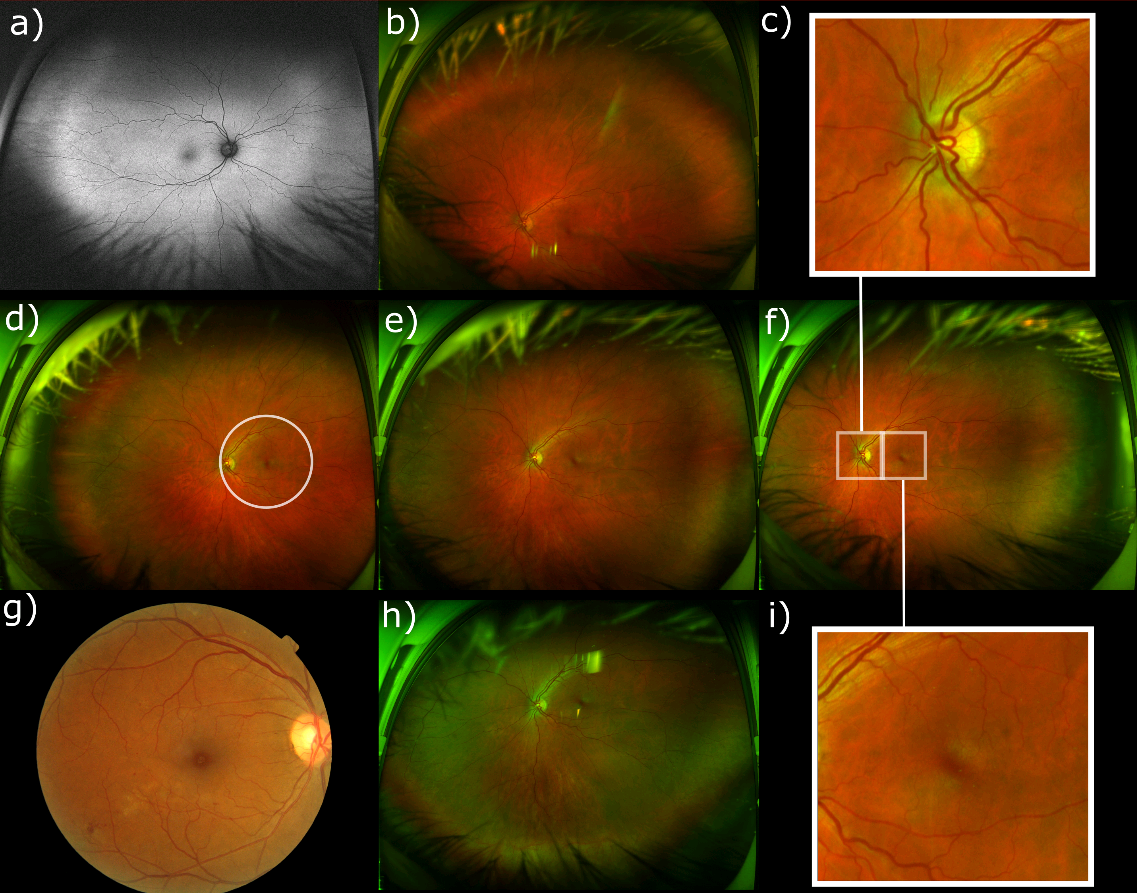}}
{\caption{\label{multipanel}{Example \ac{UWFoV-SLO} and fundus camera (FC) images; a) Right eye autofluorescence (AF) central pole gaze (CP), b) Left eye red-green (RG) reflectance eyesteered gaze (ES) Superior c) Left eye optic disc d) Left eye RG ES Nasal (approximate FoV of FC superimposed) e) Left eye RG CP f) Left eye RG ES Temporal g) Right eye FC \cite{Meyergithub} h) Left eye RG ES Inferior i) Left eye Fovea. Images b), d):f) and h) make a complete set of CP and ES images of a left eye.}}}
\end{figure}

\makeatletter
\AC@reset{UWFoV-SLO}
\AC@reset{AF}
\AC@reset{RG}
\AC@reset{FC}
\AC@reset{CNN}
\makeatother

The automatic classification of the \emph{laterality} of the image, i.e., whether the image is of the right or left eye, is of clinical interest as it allows the image to be labeled without manual input from an operator. This can save valuable time in the clinical environment, with the benefit of classification accuracy similar to  human grading \cite{Jang2018}, as operators are prone to error due to the repetitive nature of this task. The laterality  of an image can be inferred from the coordinates of the \ac{OD} and fovea; in images of the left eye, the \ac{OD} is to the left of the fovea. The opposite is true for the right eye (see Fig. \ref{multipanel}).

A fundus camera captures white-light photographs of the retina, whereas an \ac{SLO} uses directed laser beams to measure reflectance at each point on the retina to build a reflectance image. This technique allows imaging of a wider field of view --- for instance, Optos \acp{UWFoV-SLO} used in this study attain $200 \degree$ \ac{FoV}, covering 82\% of the retina. Fundus camera images cover a smaller \ac{FoV} (typically $45 \degree$), resulting in features such as the \ac{OD} appearing proportionally larger with respect to the imaged area in fundus camera images compared to \acp{UWFoV-SLO} images. See Fig. \ref{multipanel} d) and \ref{multipanel} g) for a \ac{FoV} comparison between fundus camera images and \ac{UWFoV-SLO} images.

In addition to \ac{RG} reflectance images, \ac{UWFoV-SLO} systems are capable of capturing \ac{AF} images, in which a laser is used to stimulate emission at longer wavelengths. This allows cell respiration to be visualised, and aids the detection and diagnosis of retinal diseases \cite{Holz2007}. \ac{AF} images have higher intensity levels than reflectance images, and exhibit higher levels of noise. Fig. \ref{multipanel} shows examples of \ac{RG} and \ac{AF} images for comparison.

The work presented here is novel for a number of reasons. Firstly, to the best of our knowledge this is the first time an \ac{OD} and fovea localisation method has been reported on \ac{UWFoV-SLO} images. Secondly, we are the first to investigate images captured in the \ac{AF} modality. Thirdly, we show that the proposed \ac{CNN} architecture can be trained and tested with retinal images captured in multiple modalities. Lastly, we show that the accuracy of the laterality inferred from the \ac{OD} and fovea coordinates predicted by the proposed method is higher than that of a second \ac{CNN} which was specifically trained to perform laterality classification.

\section{Related work}

Sinthanayothin et al. \cite{Sinthanayothin1999} were the first to propose an automated method to estimate the locations of the \ac{OD} and fovea in digital fundus camera photographs. The \ac{OD} was located by finding the region of highest intensity variation to the surrounding pixels, and the fovea located by finding the darkest region near the \ac{OD}. Hoover et al. \cite{Hoover2003} showed that the direction of the retinal vasculature could be used to locate the \ac{OD}, as vasculature extends from there.

Foracchia et al. \cite{Foracchia2004} used the technique of fitting parametric models to the blood vessels to locate the \ac{OD} after extracting the vasculature. Fleming et al. \cite{Fleming2007} showed that the retinal vasculature could be modeled as an elliptical form and analysed with the Hough transform, yielding approximate locations of the \ac{OD} and fovea. These locations were refined with edge and intensity measurements. Tobin et al. \cite{Tobin2007} took the segmented retinal vasculature map, and spatially measured parameters such as vessel density and thickness. This was used with a geometric model of the retina for \ac{OD} and fovea location estimation.

Niemeijer et al. \cite{Niemeijer2009} was first to treat retinal landmark localisation as a regression problem, in which the landmark coordinates were to be estimated. They showed that parameters estimated from the vasculature could be used to train a kNN regressor to return the coordinates of both the \ac{OD} and fovea. Morphological operations on the blood vessels have also been used to determine the location of the \ac{OD}, for example the work of Marin et al. \cite{Marin2015}.

With the advent of deep learning, methods employing these have been developed to locate retinal landmarks in fundus camera images. Calimeri et al. \cite{Calimeri2016} employed transfer-learning of a \ac{CNN} trained for face-detection for \ac{OD} detection.  Niu et al. \cite{Niu2017} showed the region in the image containing the \ac{OD} could be determined by generating saliency maps from the image (maps which indicate `interesting' features, such as high spatial frequency). From these maps, a number of candidate \ac{OD} regions were selected. These were then processed with a \ac{CNN}, which was trained to classify each region as containing \ac{OD} or not. Similarly, Mitra et al. \cite{Mitra2018} used a \ac{CNN} to return a region-of-interest containing the \ac{OD}, and Meng et al. \cite{Meng2018} used a \ac{CNN} trained on images where the blue channel is replaced with a vasculature map, to locate the \ac{OD}.

Simultaneous \ac{OD} and fovea localisation in fundus camera images with \acp{CNN} has been shown by Al-Bander et al. \cite{Al-Bander2016}, whereby a single \ac{CNN} returned four outputs, the $x$ and $y$ coordinates of the \ac{OD} and fovea. This was further developed \cite{Al-Bander2018}, in which two networks were utilised --- the first gave an approximate location of the coordinates of both features, which were used to extract two image patches centred around the \ac{OD} and fovea respectively. These image patches were then passed to the second network for refined feature centre estimation.

Recently, Meyer et al. \cite{Meyer2018} showed that a U-net architecture \cite{unet} could be used to predict the distance of each pixel from the nearest landmark. A predicted distance map was calculated, and image processing techniques were employed to identify the two landmarks in the map. At this stage, it was not known which landmark was the \ac{OD} and which the fovea. A further step, based on assumptions of retinal brightness, was used to label the landmarks.

Automated laterality determination using fundus camera images has also been studied. For example Tan et al. \cite{Tan2010} extracted blood vessels and the \ac{OD} which were passed to a support vector machine for laterality classification. Roy et al. \cite{Roy2016} deployed transfer-learning with a \ac{CNN}, which also contained auxiliary inputs for extracted image data, such as blood vessel density and orientation. More recently, Jang et al. \cite{Jang2018} demonstrated that \acp{CNN} could be put to this task, achieving 99.0\% classification accuracy.

\section{Materials}

Images captured in both \ac{RG} and \ac{AF} modes were used in this experiment. The majority of images were captured in the \ac{CP} gaze, in which the scan is centred on the fovea. Images were also captured with the \ac{ES} gazes, in which the scan was centred on the superior, inferior, nasal or temporal fields of the retina (Fig. \ref{multipanel}). The data set contained 4732 left eye images and 4786 right eye images. The data set was doubled in size by generating a horizontally-flipped version of each image, and the associated laterality class was swapped.

Table \ref{numbertab} shows the number of images and subjects from the \ac{RG} and \ac{AF} modes, split by gaze (CP or ES), including the horizontally flipped version of each original image. Some subjects were imaged in CP and ES or RG and AF so subject totals are not the sum of rows and columns.

\begin{table}
  \begin{center}
    \caption{Number of images, including horizontally flipped versions, split by modality and gaze. Number of subjects indicated in parentheses.}\label{numbertab}
\begingroup
\setlength{\tabcolsep}{10pt} % Default value: 6pt
\renewcommand{\arraystretch}{1.1} % Default value: 1
\begin{tabular}{|l|rrr|}
\hline
\backslashbox{Mode}{Gaze} &  \multicolumn{1}{c}{CP}  & \multicolumn{1}{c}{ES} & \multicolumn{1}{c|}{Total}\\
\hline
RG    & 11132 (1153) & 1818  \,\,\:(87) & 12950 \, (1157)\\
AF    & 5792 \: (987) & 294  \,\,\:(77) & 6086  \,\,\,\: (993)\\
Total & 16924 (2139)& 2112 (164) & \textbf{19036 (2149)}\\
\hline

\end{tabular}
\endgroup
\end{center}
\end{table}

Annotations of all images in the data set were obtained from a trained observer, who was required to annotate the location of the \ac{OD} and fovea, and the image laterality and gaze. Three additional graders were required to annotate 100 RG and 100 AF images, so that inter-grader variation could be assessed.

\section{Method}
A CNN was devised, which accepted an \ac{UWFoV-SLO} retinal image as input, and predicted four parameters; the OD and fovea $x$ and $y$ coordinates.

\subsection{Preprocessing}
To avoid overfitting and obtain a fair evaluation of results, images were split into training, validation and test sets on the constraints that 1) all images from any one subject were not put into more than one of the training, validation or test sets 2) The training, validation or test sets contained approximately 70\%, 20\% and 10\% of the data set, respectively.

Images were downsampled from 3072$\times$3900 to 768$\times$975 pixels to reduce computational complexity. Only the green reflectance channel was used, so a single network could be devised to accept both reflectance and \ac{AF} images, which are composed of one channel. The green reflectance channel was selected because the vasculature (the topography of which can be used for inferring laterality) contrast was higher. Images were not cropped, as landmarks can be located in the extremes of the capture area in \ac{ES} images (Fig. \ref{multipanel}). Image intensity values were scaled to have zero mean and unit standard deviation.

\ac{OD} and fovea coordinates were normalised by dividing both $x$ and $y$ coordinates by the image width, $W$, so all coordinates were in the range [0,1], and the loss function was not biased in favour of the $y$ coordinate. Predicted values were mapped back to image coordinates at the inference stage. In the case of horizontally flipped images, the flipped $x$ coordinates of the \ac{OD} and fovea, $x'$, were calculated by $x' = W - x$.

\subsection{Architecture}
\begin{figure}[htbp]

\includegraphics[width=\linewidth]{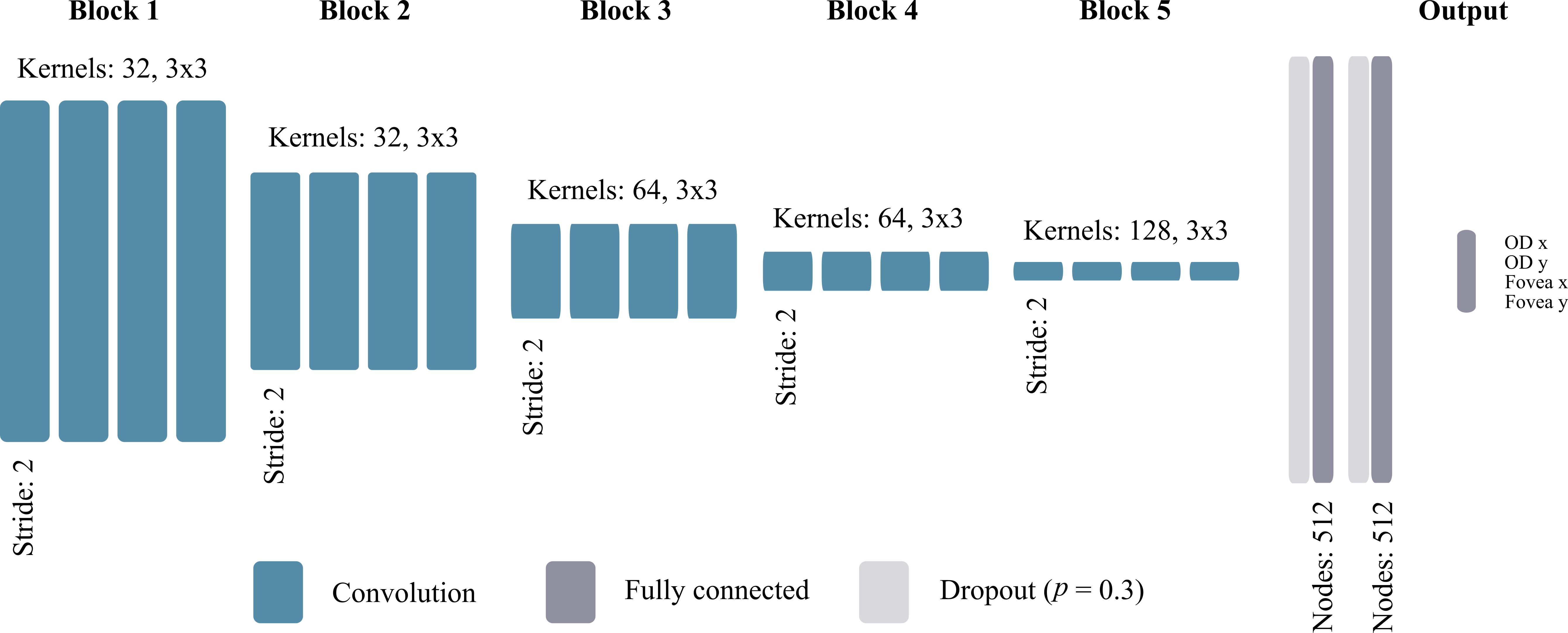}
\caption{\label{arch}{Illustration of the proposed architecture for landmark localisation.}}
\end{figure}

Five blocks of four convolutional layers were stacked, and convolutional layers in the same block had the same number of kernels. Blocks 1 and 2 had 32 kernels, 3 and 4 had 64 kernels, 5 had 128 kernels. The initial convolutional layer of each block had a stride of 2, all subsequent convolutional layers in each block had stride 1. The output from the final convolutional block was flattened, and two fully connected layers of 512 nodes followed, each preceded by dropout layers of dropout probability $p=0.3$ \cite{dropout}. The output layer was a fully connected layer of four nodes, one for each $x$ and $y$ coordinate of \ac{OD} and fovea (Fig. \ref{arch}).

The network was optimised using the Adam optimiser \cite{adam} with learning rate of 0.0001 and decay rate parameters $\beta_1 = 0.9$ and $\beta_2 = 0.999$. The norm of the gradients was clipped at 1.0 to avoid instability of the loss during training. The mean squared error loss function between the ground truth landmark coordinates and the predicted coordinates was used.

The network (containing 49,882,660 trainable parameters) was trained using Keras with TensorFlow backend, and was performed on a mixture of \ac{RG} and \ac{AF} images presented in batches of 16 images. 13,550 training images from 1,587 subjects were used.

\subsection{Landmark localisation evaluation}
\ac{OD} and fovea localisation accuracy was defined as the percentage of images for which the Euclidean distance between the predicted landmark location and the ground truth was below $n$ \ac{OD} radii ($r$).

To determine the \ac{OD} radius, a modification to the \ac{OD} diameter approximation proposed by Al-Bander et al. was used \cite{Al-Bander2018}. The \ac{OD} radius $r$ for each image was estimated from the Euclidean distance between the ground truth $x$ and $y$ coordinates of the \ac{OD} and fovea (Equation \ref{rho}) as

\begin{equation}
  \label{rho}
  r = \frac{\sqrt{ (x_{\text{OD}}- x_{\text{fovea}})^2 + (y_{\text{OD}}- y_{\text{fovea}})^2  }}{5}
\end{equation}

The OD and fovea annotations of three additional graders for 100 RG and 100 AF images were compared to the ground-truth labels. The mean Euclidean distance and the between the annotations, and the standard deviation, was calculated.

An open-source implementation \cite{Meyergithub} of the method for \ac{OD} and fovea localisation in fundus camera images proposed by Meyer et al. \cite{Meyer2018} was tested on \ac{RG} and \ac{AF} \ac{UWFoV-SLO} images. Given the constraints on the input size of the pre-trained network, \ac{RG} and \ac{AF} images were resized to $384\times512$ pixels, and padded with zeros to $512\times512$ pixels. Distance probability maps were analysed with the image processing pipeline presented \cite{Meyergithub} to predict the \ac{OD} and fovea coordinates.

\subsection{Laterality classification evaluation}
Laterality accuracy was calculated as the percentage of images which were classified as belonging to the correct class, from left or right eye. The class sizes were balanced as a result of the flipping operation, thus the accuracy score did not require weighting.

The laterality of the input image can be inferred from the OD and fovea $x$ coordinates, as shown in Equation \ref{lat}.

\begin{equation}
  \label{lat}
  \text{Laterality}=\begin{cases}
    \text{L}, & \text{if $x_{\text{OD}}<x_{\text{fovea}}$}.\\
    \text{R}, & \text{if $x_{\text{OD}}>x_{\text{fovea}}$}.
  \end{cases}
\end{equation}

A modified version of the proposed architecture was used to implement a laterality classifier. The output layer was replaced with a fully connected layer with a single, Sigmoid activated node to predicted the laterality class of the input image. This allowed the laterality inferred from the landmark coordinates predicted by the proposed method to be compared with a classifier baseline. The classification network was trained in the same fashion as the landmark localisation network, but using binary cross-entropy loss between the ground truth class and predicted class labels.

\section{Results}
\begin{figure}[htbp]
\begin{centering}
  \hspace*{1.3cm}
\includegraphics[width=0.85\linewidth]{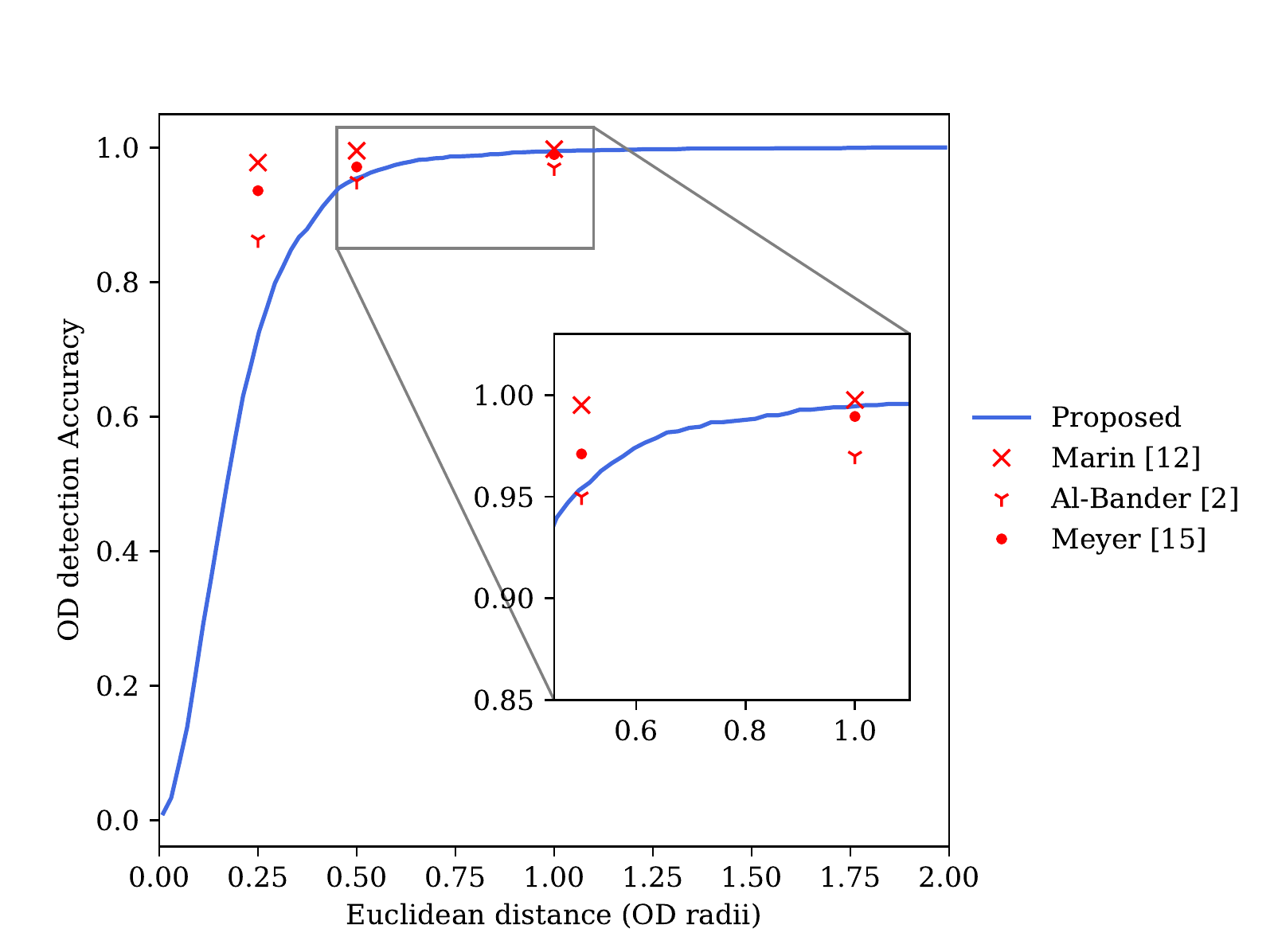}
\caption{\label{odacc}Optic disc localisation accuracy for the proposed method as a function of Euclidean distance, expressed in optic disc radii. Results of other methods, achieved on fundus camera datasets, are shown for comparison.}
\end{centering}
\end{figure}

\begin{figure}[htbp]
\begin{centering}
  \hspace*{1.3cm}
\includegraphics[width=0.85\linewidth]{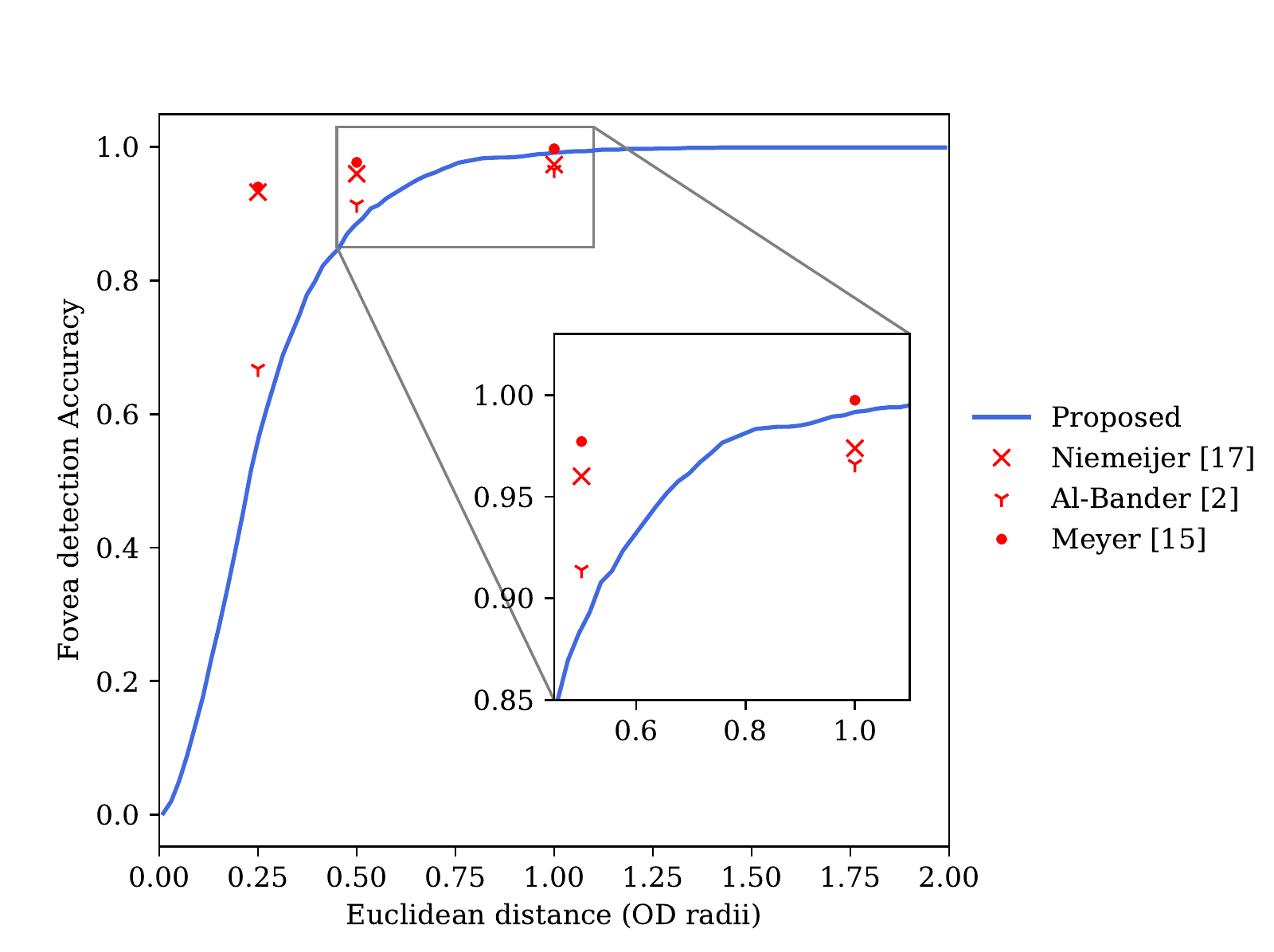}
\caption{\label{fovacc}Fovea localisation accuracy for the proposed method as a function of Euclidean distance, expressed in optic disc radii. Results of other methods, achieved on fundus camera datasets, are shown for comparison.}
\end{centering}
\end{figure}

\FloatBarrier
Results for landmark localisation accuracy and laterality classification accuracy for 1790 test images are shown in Table \ref{acctab}. Results of the proposed method are split by modality and gaze. Other methods demonstrated on fundus camera images are presented for comparison. Plots of \ac{OD} and fovea accuracy as as function of distance are shown in Fig. \ref{odacc} and Fig. \ref{fovacc}. The accuracy of laterality inferred from the $x$ coordinates of the \ac{OD} and fovea is shown in Table \ref{acctab}. The laterality classification network accuracy is also shown.

%$r$
\begin{table}[htbp]
  \centering
  \caption{\label{acctab} Accuracy for landmark localisation (expressed in OD radii, $r$), and accuracy of laterality classification inferred (Inf.) from the landmark $x$ coordinates, and classified by classifier network (Class.). Results are split by mode and gaze. Accuracies are expressed as percentages. Fundus camera (FC) methods shown for comparison.}
    \begin{tabular}{l|ll|rrr|rrr|rr|r}
    \textbf{Method} & \textbf{Mode} & \textbf{Gaze} & \multicolumn{3}{c|}{\textbf{OD Accuracy}} & \multicolumn{3}{c|}{\textbf{Fovea accuracy}} & \multicolumn{2}{c|}{\textbf{Lat. Acc.}} & \multicolumn{1}{l}{\textbf{Images}} \\
    \hline
          &       &       & \multicolumn{1}{c}{\textbf{0.25$r$}} & \multicolumn{1}{c}{\textbf{0.5$r$}} & \multicolumn{1}{c|}{\textbf{1$r$}} & \multicolumn{1}{c}{\textbf{0.25$r$}} & \multicolumn{1}{c}{\textbf{0.5$r$}} & \multicolumn{1}{c|}{\textbf{1$r$}} & \multicolumn{1}{l}{\textbf{Inf.}} & \multicolumn{1}{l|}{\textbf{Class.}} &  \\
    Jang \cite{Jang2018} & FC    &       & \multicolumn{1}{c}{x} & \multicolumn{1}{c}{x} & \multicolumn{1}{c|}{x} & \multicolumn{1}{c}{x} & \multicolumn{1}{c}{x} & \multicolumn{1}{c|}{x} & \multicolumn{1}{c}{x} & 99.0  & 5180 \\
    Niemeijer \cite{Niemeijer2009} & FC    &       & \multicolumn{1}{c}{x} & \multicolumn{1}{c}{x} & \multicolumn{1}{c|}{x} & 93.3  & 96.0  & 97.4  & \multicolumn{1}{c}{x} & \multicolumn{1}{c|}{x} & 800 \\
    Marin \cite{Marin2015} & FC    &       & 97.8  & 99.5  & 99.8  & \multicolumn{1}{c}{x} & \multicolumn{1}{c}{x} & \multicolumn{1}{c|}{x} & \multicolumn{1}{c}{x} & \multicolumn{1}{c|}{x} & 1200 \\
    Al-Bander \cite{Al-Bander2018} & FC    &       & 83.6  & 95.0  & 97.0  & 66.8  & 91.4  & 96.6  & \multicolumn{1}{c}{x} & \multicolumn{1}{c|}{x} & 1200 \\
    Meyer \cite{Meyer2018} & FC    &       & 93.6  & 97.1  & 98.9  & 94.0  & 97.7  & 99.7  & \multicolumn{1}{c}{x} & \multicolumn{1}{c|}{x} & 1136 \\
    \hline
    Proposed & RG    & All   & 67.6  & 93.7  & 99.5  & 56.0  & 88.2  & 99.1  & 99.8  & 99.0  & 1144 \\
    Proposed & RG    & CP    & 71.3  & 95.1  & 99.5  & 58.7  & 90.4  & 99.0  & 99.8  & 98.9  & 1006 \\
    Proposed & RG    & ES    & 40.6  & 83.3  & 99.3  & 36.2  & 72.5  & 100.0 & 100.0 & 100.0 & 138 \\
    Proposed & AF    & All   & 73.4  & 96.3  & 99.2  & 48.6  & 85.3  & 99.1  & 100.0 & 100.0 & 646 \\
    Proposed & AF    & CP    & 74.9  & 97.0  & 99.3  & 49.3  & 85.9  & 99.3  & 100.0 & 100.0 & 610 \\
    Proposed & AF    & ES    & 47.2  & 83.3  & 97.2  & 36.1  & 75.0  & 94.4  & 100.0 & 100.0 & 36 \\
    \textbf{Proposed} & \textbf{All} & \textbf{All} & \textbf{69.7} & \textbf{94.6} & \textbf{99.4} & \textbf{53.4} & \textbf{87.2} & \textbf{99.1} & \textbf{99.9} & \textbf{99.4} & \textbf{1790} \\
    Proposed & All   & CP    & 72.6  & 95.9  & 99.4  & 55.2  & 88.7  & 99.1  & 99.9  & 99.3  & 1616 \\
    Proposed & All   & ES    & 42.0  & 83.3  & 98.9  & 36.2  & 73.0  & 98.9  & 100.0 & 100.0 & 174 \\
    \end{tabular}%
  \label{tab:addlabel}%
\end{table}%

The mean Euclidean distance $\mu$ and standard deviation $\sigma$, normalised by $r$, between three additional graders and the ground-truth landmark coordinates is shown in Table \ref{grader_compare}. 100 RG and 100 AF images were used. The proposed method, tested on 1790 test images, is shown for comparison.

% Table generated by Excel2LaTeX from sheet 'Sheet1'
% Table generated by Excel2LaTeX from sheet 'Sheet1'

% CONVERT MEAN TO \mu, std to \sigma
%$\mu$
%$\sigma$
%$r$
% Table generated by Excel2LaTeX from sheet 'Sheet1'
\begin{table}[htbp]
  \centering
  \caption{\label{grader_compare} Mean Euclidean distance $\mu$ and standard deviation $\sigma$ between three graders and the ground-truth landmark coordinates for 100 RG and 100 AF images. Distances normalised by $r$. The mean of grader scores, as well as the proposed method when tested on 1790 test images, are shown for comparison.}
    \begin{tabular}{l|rrrr|rrrr}
          & \multicolumn{4}{c|}{\textbf{RG}} & \multicolumn{4}{c}{\textbf{AF}} \\
    \textbf{Grader} & \multicolumn{1}{l}{\textbf{OD $\mu$}} & \multicolumn{1}{l}{\textbf{OD $\sigma$}} & \multicolumn{1}{l}{\textbf{Fovea $\mu$}} & \multicolumn{1}{l|}{\textbf{Fovea $\sigma$}} & \multicolumn{1}{l}{\textbf{OD $\mu$}} & \multicolumn{1}{l}{\textbf{OD $\sigma$}} & \multicolumn{1}{l}{\textbf{Fovea $\mu$}} & \multicolumn{1}{l}{\textbf{Fovea $\sigma$}} \\
    \hline
    \multicolumn{1}{l|}{1} & 0.127 & 0.074 & 0.205 & 0.159 & 0.117 & 0.073 & 0.323 & 0.222 \\
    \multicolumn{1}{l|}{2} & 0.126 & 0.082 & 0.167 & 0.112 & 0.139 & 0.078 & 0.318 & 0.238 \\
    \multicolumn{1}{l|}{3} & 0.164 & 0.089 & 0.198 & 0.120 & 0.139 & 0.093 & 0.331 & 0.239 \\
    \multicolumn{1}{l|}{Mean} & 0.139 & 0.082 & 0.190 & 0.130 & 0.132 & 0.081 & 0.324 & 0.233 \\
    \hline
    \textbf{Proposed} & \textbf{0.226} & \textbf{0.174} & \textbf{0.278} & \textbf{0.219} & \textbf{0.207} & \textbf{0.160} & \textbf{0.302} & \textbf{0.205} \\
    \end{tabular}%
  \label{tab:addlabel}%
\end{table}%

The pixel landmark distance prediction method proposed by Meyer \cite{Meyer2018} achieved 0\% accuracy for both \ac{OD} and fovea localisation within $1r$ on 1790 \ac{UWFoV-SLO} images. The inferred laterality classification accuracy was 52.7\%. These results could be expected given the model was trained on fundus camera images, and highlight the difference between the two image capturing systems.

\FloatBarrier

\section{Conclusion}
In this paper, we have proposed a \ac{CNN} architecture for the localisation of the \ac{OD} and the fovea in \ac{UWFoV-SLO} images of two modalities --- \ac{RG} reflectance and \ac{AF} emission. The method achieved an \ac{OD} localisation accuracy of 99.4\% within one \ac{OD} radius, and fovea localisation accuracy of 99.1\% within one \ac{OD} radius (Table \ref{acctab}). These values are higher than those from other methods applied to fundus camera images, found in the literature, at 0.5$r$ and 1$r$ distances, but not at 0.25$r$. However, a fair, direct comparison between these results cannot be made due to the differences between the imaging techniques. The \ac{UWFoV-SLO} has a larger \ac{FoV}, which results in features such as the \ac{OD} and fovea appearing smaller compared to fundus camera images (Fig. \ref{multipanel}). When comparing the performance with human graders, the proposed method achieves AF fovea localisation distance errors comparable with inter-grader variation (Table \ref{grader_compare}).

We have shown that this CNN can also be used to infer the laterality of the input image by considering the predicted $x$ coordinates of the retinal landmarks. This method achieves laterality classification accuracy values higher than a similar network trained only to predict the laterality of the input image (99.9\% vs 99.4\%), and results are in line with published literature using fundus camera images \cite{Jang2018}.

Testing of a \ac{CNN} with weights trained on fundus camera images \cite{Meyergithub,Meyer2018}  on \ac{UWFoV-SLO} images lead to extremely low accuracy values. This highlights the inherent differences between the images captured by fundus cameras and \acp{UWFoV-SLO}, and confirms that direct transfer of pre-trained networks between the two image types is not possible.

Future work will involve retraining the Meyer model on \ac{UWFoV-SLO} images to compare performance. However, modification to the image processing technique for determining which landmark is the \ac{OD} and fovea will be required, as the existing method assumes that the \ac{OD} is brighter than the fovea, which does not hold in the case of \ac{AF} images. Future work will also include training and testing our method on fundus camera data sets, ideally of subjects also imaged with \acp{UWFoV-SLO}, so direct comparison can be made.

% ---- Bibliography ----
%
% BibTeX users should specify bibliography style 'splncs04'.
% References will then be sorted and formatted in the correct style.
%

\bibliography{LateralityODfov}

\end{document}